\documentclass[nofootinbib,prd,twocolumn,showpacs,showkeys,preprintnumbers]{revtex4}
\usepackage{hyperref,amssymb,amsmath,mathrsfs,bm,graphicx}
\begin{document}
\title{ON THE ROLE OF ELECTRIC CHARGE AND COSMOLOGICAL CONSTANT IN STRUCTURE SCALARS}
\author{L. Herrera \footnote{Also at U.C.V., Caracas}} 
\email{laherrera@cantv.net.ve}
\affiliation{Departamento   de F\'{\i}sica Te\'orica e Historia de la  Ciencia,
Universidad del Pa\'{\i}s Vasco, Bilbao, Spain}
\author{A. Di Prisco \footnote{Also at U.C.V., Caracas}}
\email{adiprisc@fisica.ciens.ucv.ve}
\affiliation{Departamento   de F\'{\i}sica Te\'orica e Historia de la  Ciencia,
Universidad del Pa\'{\i}s Vasco, Bilbao, Spain}
\author{J. Ib\'a\~nez } 
\email{j.ibanez@ehu.es}
\affiliation{Departamento   de F\'{\i}sica Te\'orica e Historia de la  Ciencia,
Universidad del Pa\'{\i}s Vasco, Bilbao, Spain}
\date{\today}
\begin{abstract}
The physical meaning of structure scalars is analyzed for charged dissipative spherical fluids and for neutral dust in the presence of cosmological constant. The role played by such factors in the structure scalars is clearly brought out and physical consequences are discussed. Particular attention needs to be paid to the changes introduced by the above mentioned factors in the inhomogeneity factor and the evolution of the expansion scalar and the shear tensor.
\end{abstract}
\date{\today}
\maketitle
\section{INTRODUCTION}
In a recent work \cite{H1}, the full set of equations governing the structure and the  evolution of self--gravitating spherically symmetric dissipative fluids with anisotropic stresses, was written down in terms of  five scalar quantities obtained from the orthogonal splitting of the Riemann tensor   in the context of general relativity.  It was shown that these scalars (denoted by $X_T$, $X_{TF}$, $Y_{TF}$,  $Y_T$   and $Z$) are directly related to fundamental properties of the fluid distribution, such as: energy density, energy density inhomogeneity, local anisotropy of pressure, dissipative flux and the active gravitational mass. In particular the following properties of such quantities were established:
\begin{itemize}
\item $X_T$  is the energy--density of the fluid whereas  $Z$ describes all possible dissipative fluxes \cite{H1}. 
\item  In the absence of dissipation,   $X_{TF}$ controls inhomogeneities in the energy density \cite{H1}. 
\item $Y_{TF}$ describes the influence of the local anisotropy of pressure and density inhomogeneity on the Tolman mass \cite{H1}.
\item $Y_T$ turns out  to be proportional to the  Tolman mass ``density'' for systems in equilibrium or quasi--equilibrium \cite{H1}.
\item The evolution of the expansion scalar and the shear tensor  is fully controlled by $Y_{TF}$ and $Y_T$ \cite{H1, H2, H3}.
\end{itemize}

Motivated by the deep physical meaning of structure scalars we shall in this work calculate them for two situations of evident  physical interest (see for example \cite{H4, en1, e1, e2, e3, e4,  l2, l1} and references therein), namely:
\begin{itemize}
\item charged fluids.
\item neutral dust with cosmological constant.
\end{itemize}

As we shall see here both factors (electric charge and cosmological constant)  affect the evolution of the system exclusively through their presence in some of the structure scalars, stressing further their relevance in the study of self--gravitating systems.

\section{GENERAL EQUATIONS AND DEFINITIONS}
Full details of some intermediate calculations, notation and basic equations can be found in \cite{H1, H2, H3, H4}, however for self--consistency we shall here provide  a summary of the more essential equations and definitions.

We shall consider a general spherically symmetric  line element of the form

\begin{equation}
ds^2=-A^2dt^2+B^2dr^2+R^2(d\theta^2+\sin^2\theta d\phi^2),
\label{1}
\end{equation}
and a general fluid distribution whose energy--momentum tensor  may be written as
\begin{eqnarray}
T_{\alpha\beta}&=&(\mu +
P_{\perp})V_{\alpha}V_{\beta}+P_{\perp}g_{\alpha\beta}+(P_r-P_{\perp})\chi_{
\alpha}\chi_{\beta}\nonumber \\&+&q_{\alpha}V_{\beta}+V_{\alpha}q_{\beta}+
\epsilon l_{\alpha}l_{\beta}-2\eta\sigma_{\alpha\beta}, 
\label{3}
\end{eqnarray}
where $\mu$ is the energy density, $P_r$ the radial pressure,
$P_{\perp}$ the tangential pressure, $q^{\alpha}$ the heat flux,
$\epsilon$ the radiation density, $\eta$ the coefficient of
shear viscosity, $\sigma_{\alpha \beta}$  the shear tensor,  $V^{\alpha}$ the four velocity of the fluid,
$\chi^{\alpha}$ a unit four vector along the radial direction
and $l^{\alpha}$ a radial null four vector. The four--vectors above
for  (\ref{1}) are 
\begin{eqnarray}
V^{\alpha}=A^{-1}\delta_0^{\alpha}, \;\;
q^{\alpha}=qB^{-1}\delta^{\alpha}_1, \;\;\nonumber \\
l^{\alpha}=A^{-1}\delta^{\alpha}_0+B^{-1}\delta^{\alpha}_1, \;\;
\chi^{\alpha}=B^{-1}\delta^{\alpha}_1, \label{5}
\end{eqnarray}
where  $q$ is a function of  $t$ and $r$, and   $q^\alpha = q
\chi^\alpha$.

If the fluid is charged we shall need to add the electromagnetic contribution to the fluid distribution.

The electromagnetic energy tensor $S_{\alpha\beta}$ is given by (see \cite{H4} for details)
\begin{equation}
S_{\alpha\beta}=\frac{1}{4\pi}\left({F_{\alpha}}^{\gamma}F_{\beta\gamma}
-\frac{1}{4}F^{\gamma\delta}F_{\gamma\delta}g_{\alpha\beta}\right),
\label{6}
\end{equation}
where $F_{\alpha\beta}$ is the electromagnetic field tensor.
The electric charge interior to radius $r$ is time independent, and given by 
\begin{equation}
s(r)=4\pi\int^r_0\varsigma BR^2dr, \label{13}
\end{equation}
where the charge density  $\varsigma$, is a function
of $t$ and $r$. 

Next, for the four--acceleration, the expansion scalar and the shear tensor we have
\begin{equation}
a_1=\frac{A^{\prime}}{A}, \;\;
a^2=a^{\alpha}a_{\alpha}=\left(\frac{A^{\prime}}{AB}\right)^2,
\label{5c}
\end{equation}
with  $a^\alpha= a \chi^\alpha$,
\begin{equation}
\Theta={V^{\alpha}}_{;\alpha}=\frac{1}{A}\left(\frac{\dot{B}}{B}+2\frac{\dot{R}}{R}\right),
\label{th}
\end{equation}
and
\begin{equation}
\sigma_{11}=\frac{2}{3}B^2\sigma, \;\;
\sigma_{22}=\frac{\sigma_{33}}{\sin^2\theta}=-\frac{1}{3}R^2\sigma,
 \label{5a}
\end{equation}
with
\begin{equation}
\sigma^2=\frac{3}{2}\sigma^{\alpha\beta}\sigma_{\alpha\beta}=\frac{1}{A^2}\left(\frac{\dot{B}}{B}-\frac{\dot{R}}{R}\right)^2,
\label{5b}
\end{equation}
where dots and primes denote differentiation with respect to $t$ and $r$ respectively.

The mass function $m(t,r)$ 
is given by 
\begin{equation}
m=\frac{(R)^3}{2}{R_{23}}^{23} + \frac{s^2}{2R}
=\frac{R}{2}\left[\left(\frac{\dot{R}}{A}\right)^2
-\left(\frac{R^{\prime}}{B}\right)^2+1\right],
 \label{18}
\end{equation}
which  can be rewritten as
\begin{equation}
E \equiv \frac{R^{\prime}}{B}=\left(1+U^2-\frac{2m(t,r)}{R} + \frac{s^2}{R^2}\right)^{1/2},
\label{20}
\end{equation}
where  $U$ is the areal velocity of the collapsing
fluid i.e. $U=\frac{1}{A}\dot R$.

From 
(\ref{18})
 we may obtain (see (38) in \cite{H4})
\begin{equation}
m=\int^{r}_{0}4\pi R^2 \left(\tilde \mu + \tilde q \frac{U}{E}\right)R^{\prime}dr + \frac{s^2}{2R} + \frac{1}{2} \int^r_0{\frac{s^2}{R^2}R^{\prime}dr}.\label{27int}
\end{equation}
(assuming a regular centre to the distribution, so $m(0)=0$), 
or
\begin{eqnarray}
\frac{3m}{R^3} &=& 4 \pi \tilde \mu - \frac{4 \pi}{R^3}\int^r_0{R^3  \tilde \mu^{\prime} dr} + \frac{4 \pi}{R^3}\int^r_0{3 \tilde q UBR^2 dr}\nonumber\\
&+&\frac{3}{R^3}\left(\frac{s^2}{2R} + \frac{1}{2} \int^r_0{\frac{s^2}{R^2} R^{\prime}dr}\right),
\label{3mi}
\end{eqnarray}
where $\tilde \mu = \mu + \epsilon$ and $\tilde q = q + \epsilon$.

The Weyl tensor ($ C_{\alpha \mu \beta \nu}$) as usual may be decomposed in its electric and magnetic parts, however due to the spherical symmetry, the magnetic part vanishes and so the Weyl tensor is expressed in terms of its electric part alone.

The electric  part of  Weyl tensor is defined by 
\begin{equation}
E_{\alpha \beta} = C_{\alpha \mu \beta \nu} V^\mu V^\nu,
\label{elec}
\end{equation}
which  may also be  writen as:
\begin{equation}
E_{\alpha \beta}={\cal E} (\chi_\alpha
\chi_\beta-\frac{1}{3}h_{\alpha \beta}), \label{52}
\end{equation}
where
\begin{equation}
h_{\alpha \beta}=g_{\alpha \beta}+V_\alpha V_\beta,
\label{nh}
\end{equation}
and 
\begin{widetext}
\begin{eqnarray}
{\cal E}= \frac{1}{2 A^2}\left[\frac{\ddot R}{R} - \frac{\ddot B}{B} - \left(\frac{\dot R}{R} - \frac{\dot B}{B}\right)\left(\frac{\dot A}{A} + \frac{\dot R}{R}\right)\right]
+ \frac{1}{2 B^2} \left[\frac{A^{\prime\prime}}{A} -
\frac{R^{\prime\prime}}{R} + \left(\frac{B^{\prime}}{B} +
\frac{R^{\prime}}{R}\right)\left(\frac{R^{\prime}}{R}-\frac{A^{\prime}}{A}\right)\right]
- \frac{1}{2 R^2}. \label{E}
\end{eqnarray}
\end{widetext}

Using Einstein equations, (\ref{18}) (\ref{3mi}) and (\ref{E}) we can write
\begin{widetext}
\begin{equation}
 {\cal E}= 4\pi \left(2\eta \sigma-\Pi \right) + \frac{3 s^2}{2R^4}+ \frac{4 \pi}{R^3}\int^r_0{R^3  \tilde \mu^{\prime} dr}- \frac{12 \pi}{R^3}\int^r_0{ \tilde q UBR^2 dr}-\frac{3}{2R^3}\int^r_0\frac{s^2R^{\prime}}{R^2}dr,
\label{mW}
\end{equation}
\end{widetext}
where $\Pi=\tilde P_{r}-P_{\bot}$ and $\tilde P_{r}=P_r+\epsilon$.

\section{STRUCTURE SCALARS FOR THE CHARGED FLUID}
We can now calculate the structure scalars for our charged fluid. For doing that 
let us define tensors $Y_{\alpha \beta}$ and 
$X_{\alpha \beta}$ by:
\begin{equation}
Y_{\alpha \beta}=R_{\alpha \gamma \beta \delta}V^\gamma V^\delta,
\label{electric}
\end{equation}
\begin{equation}
X_{\alpha \beta}=^*R^{*}_{\alpha \gamma \beta \delta}V^\gamma
V^\delta=\frac{1}{2}\eta_{\alpha\gamma}^{\quad \epsilon
\rho}R^{*}_{\epsilon \rho\beta\delta}V^\gamma V^\delta,
\label{magnetic}
\end{equation}
where $R^*_{\alpha \beta \gamma \delta}=\frac{1}{2}\eta
_{\epsilon \rho \gamma \delta} R_{\alpha \beta}^{\quad \epsilon
\rho}$.

Tensors $Y_{\alpha \beta}$ and  $X_{\alpha \beta}$ may be expressed as 
\begin{eqnarray}
Y_{\alpha\beta}=\frac{1}{3}Y_T h_{\alpha
\beta}+Y_{TF}(\chi_{\alpha} \chi_{\beta}-\frac{1}{3}h_{\alpha
\beta}),\label{electric'}
\\
X_{\alpha\beta}=\frac{1}{3}X_T h_{\alpha
\beta}+X_{TF}(\chi_{\alpha} \chi_{\beta}-\frac{1}{3}h_{\alpha
\beta}).\label{magnetic'}
\end{eqnarray}

Then after lengthy but simple calculations, using field equations (see (20)--(23) in \cite{H4})  and (\ref{E}) we obtain
\begin{equation}
Y_T=4\pi(\tilde \mu+3\tilde P_r-2\Pi) + \frac{s^2}{R^4}, \qquad
Y_{TF}={\cal E}-4\pi(\Pi-2\eta\sigma) + \frac{s^2}{R^4},\label{EY}
\end{equation}
\begin{equation}
X_T=8\pi\tilde \mu + \frac{s^2}{R^4} , \qquad
X_{TF}=-{\cal E}-4\pi(\Pi-2\eta\sigma)+ \frac{s^2}{R^4}.\label{EX}
\end{equation}

Using  (\ref{mW}) and  (\ref{EY})    we may write $Y_{TF}$ as
\begin{eqnarray}
Y_{TF}= - 8 \pi \Pi + 16 \pi \eta \sigma + \frac{5 s^2}{2 R^4} - \frac{3}{2 R^3}\int^r_0{\frac{s^2}{R^2}R^{\prime}dr}\nonumber \\+ \frac{4 \pi}{R^3}\int^r_0{R^3\left( \tilde \mu^{\prime} - \frac{3 \tilde q BU}{R}\right)dr}.
\label{Yi}
\end{eqnarray}

At this point it would be useful to introduce the following ``effective'' variables:
\begin{equation}
-(T_{0}^{0}+S_{0}^{0})\equiv \mu_{eff} =  \tilde \mu + \frac{s^2}{8\pi R^4},
\label{mueff}
\end{equation}
\begin{equation}
T_{1}^{1}+S_{1}^{1}\equiv P^{eff}_ r=
 \left (\tilde P_r-\frac{4}{3}\eta\sigma\right)
 -\frac{s^2}{8\pi R^4},
\label{T11} 
\end{equation}
\begin{equation}
T_{2}^{2}+S^{2}_{2}\equiv P^{eff}_\bot
=\left(P_{\perp}+\frac{2}{3}\eta\sigma\right)
 +\frac{s^2}{8\pi R^4},
 \label{T22}
\end{equation}
and
\begin{equation}
P^{eff}_r-P^{eff}_{\bot}\equiv \Pi ^{eff} = \Pi -2 \eta \sigma - \frac{s^2}{4 \pi R^4}.
\label{pieff}
\end{equation}
As it is evident from the above, the effective variables are just  the corresponding ordinary variables with all contributions (from viscosity and electric charge) included. In terms of the former, the structure scalars read
\begin{equation}
Y_{TF}= - 8 \pi \Pi^{eff} + \frac{4 \pi}{R^3}\int^r_0{R^3\left( \mu_{eff}^{\prime} - \frac{3 \tilde q BU}{R}\right)dr},
\label{Yieff}
\end{equation}
\begin{equation}
X_{TF}= - \frac{4 \pi}{R^3}\int^r_0{R^3\left(\mu_{eff}^{\prime} - \frac{3 \tilde q UB}{R}\right)dr},
\label{Xieff}
\end{equation}
\begin{equation}
Y_T=4\pi(\tilde \mu_{eff}+3\tilde P_r^{eff}-2\Pi^{eff}), 
\label{scn}
\end{equation}
\begin{equation}
X_T=8\pi\tilde \mu_{eff} \label{EXn}.
\end{equation}

The remarkable fact emerging from these expressions is that the charge contribution is always absorbed into the effective variables. In the absence of electrical charge the structure scalars are obtained from (\ref{Yieff})--(\ref{EXn}), just replacing the effective variables by the corresponding ordinary ones.

In order to delve deeper into the  question about the role of electric charge in the structure and evolution of compact objects, and how this reflects in the structure scalar we shall consider three very important equations in general relativity. These are: the evolution equation for the expansion scalar (Raychaudhuri), the evolution equation for the shear \cite{H1}, \cite{H3}, \cite{60}, \cite{61} and a differential equation relating the energy density inhomogeneity with the Weyl tensor and other physical variables \cite{H1}, \cite{H5}, \cite{60}, \cite{61} .
The Raychaudhuri equation reads in our case
\begin{equation}
V^\alpha \Theta_{;\alpha}+\frac{1}{3}\Theta ^2+\frac{2}{3}\sigma
^2-a^\alpha_{;\alpha}=-Y_T, \label{c}
\end{equation}
which has exactly the same form as in the non--charged case (see (32) in \cite{H3}).
For the shear evolution equation we find
\begin{equation}
 Y_{TF}= \chi^\alpha a_{;\alpha} +a^2-\frac{a R^\prime}{B R} -V^\alpha \sigma_{;\alpha} - \frac{2}{3}\Theta \sigma - \frac{\sigma^2}{3},
\label{she}
\end{equation}
which again, has exactly the same form as in the non--charged case (see (45) in \cite{H3}).

Finally, the differential equation for the Weyl tensor and the energy density inhomogeneity can be written as 
\begin{equation}
\left(X_{TF}+4\pi \mu_{eff} \right)^\prime=  - X_{TF} \frac{3R^\prime}{R} +4 \pi \tilde q  B (\Theta - \sigma), 
\label{wpxeff}
\end{equation}
which is exactly the same expression  for the non--charged fluid, replacing the effective energy density by the energy density (see (37) in \cite{H5}).

We shall next consider the case of dust with  cosmological constant.

\section{STRUCTURE SCALARS FOR DUST WITH COSMOLOGICAL CONSTANT }
Let us consider a spherically symmetric distribution of dust with non--vanishing cosmological constant.
Then the energy--momentum tensor takes the simple form
\begin{equation}
T_{\alpha\beta}=8\pi\mu V_{\alpha}V_{\beta}, 
\label{Tl}
\end{equation}
and Einstein equations read
\begin{equation}
G_{\alpha\beta}=T_{\alpha \beta}-\Lambda g_{\alpha\beta}, \label{2l}
\end{equation}
where $\Lambda$ is the cosmological constant.

Since the fluid is obviously geodesic for our comoving observers,  we have $A^{\prime}=0$  and rescaling the time coordinate $t$, we can put $A=1$.

The mass function now can be casted into the form
\begin{equation}
m= 4\pi\int^r_0\mu\, R^2 R^{\prime}dr +\frac{\Lambda}{6}\,R^3. \label{m}
\end{equation}

From the above, the following equations may be obtained, which are the the equivalent to (\ref{3mi}) and (\ref{mW}) in the case of dust with cosmological constant,
\begin{equation}
\frac{3m}{R^3}=4 \pi \mu +\frac{\Lambda}{2} - \frac{4 \pi}{R^3}\int^r_0 \mu^{\prime} dr,
\label{DRmli}
\end{equation}
\begin{equation}
{\cal E}= \frac{4 \pi}{R^3}\int^r_0{R^3  \mu^{\prime}  dr}.
\label{Yilp}
\end{equation}

From  (\ref{E}), (\ref{electric}), (\ref{magnetic}), (\ref{electric'}) and  (\ref{magnetic'}),   with the help of Einstein equations 
we obtain for the structure scalars
\begin{equation}
Y_T=4\pi \mu - \Lambda, \qquad
Y_{TF}=-X_{TF}={\cal E}, \qquad X_T=8\pi \mu -\Lambda .
\label{EYl}
\end{equation}
Then, the evolution equations for the  shear and expansion become
\begin{equation}
{\cal E} = Y_{TF}=  - V^\alpha \sigma_{;\alpha} - \frac{2}{3}\Theta \sigma - \frac{\sigma^2}{3}, 
\label{shel}
\end{equation}
and 
\begin{equation}
V^\alpha \Theta_{;\alpha}+\frac{1}{3}\Theta ^2+\frac{2}{3}\sigma
^2-a^\alpha_{;\alpha}=-4\pi 
\mu+\Lambda=-Y_{T}, \label{c1}
\end{equation}
whereas the differential equation for the inhomogeneity factor  can be written as
\begin{equation}
\left(X_{TF}+4\pi \mu \right)^\prime= -  X_{TF} \frac{3R^\prime}{R},
\label{wpxl}
\end{equation}
from which it follows at once $\mu^{\prime}=0\leftrightarrow X_{TF}=0$, allowing us to identify $X_{TF}$ as the inhomogeneity factor.
\\
\section{SUMMARY}
In  the case of the charged fluid we have seen that the role of electrical charge in the structure and evolution of  self--gravitating systems is completley determined by structure scalars. Thus the influence of  charge, in the evolution of the expansion and the shear,  reveals itself exclusively through its contribution to $Y_T$ and $Y_{TF}$ respectively. The same can be said about the inhomogeneity factor, as it follows from (\ref{wpxeff}). It is also worth stressing the fact that the charge contribution is always absorbed into the effective variables in a rather, intuitively, obvious way.

In the case of dust with cosmological constant we see that the latter does not affect at all either the evolution of the shear or the inhomogeneity factor. Instead, it affects the evolution of the expansion scalar through the $\Lambda$ term in $Y_T$. The fact that the cosmological constant does not affect the stability of the shear--free condition deserves to be emphasized.

It should be observed that, besides local anisotropy of pressure, dissipation and shear viscosity,   the inclusion of electric charge and cosmological constant  exhausts all possible physical phenomena that we expect in a spherically symmetric relativistic fluid distribution. The fact that all of them act exclusively through their presence in structure scalars exhibits the universality of the latter.

The comments above reinforce  our belief that structure scalars are called upon to play a major role in the study of self--gravitating systems.

\begin{acknowledgments}
LH wishes to thank Fundaci\'on Empresas Polar for financial support and Departamento   de F\'{\i}sica Te\'orica e Historia de la  Ciencia, Universidad del Pa\'{\i}s Vasco, for financial support and hospitality. ADP  acknowledges hospitality of the
Departamento   de F\'{\i}sica Te\'orica e Historia de la  Ciencia,
Universidad del Pa\'{\i}s Vasco. This work was partially supported by the Spanish Ministry of Science and Innovation (grant FIS2010-15492). 
\end{acknowledgments} 

 \end{document}